\documentclass{PoS}
\usepackage{amsmath}
\usepackage{amsfonts}
\usepackage{mathrsfs}
\usepackage{placeins}

\graphicspath{{graph/}}

\title{Improving DWF Simulations: the Force Gradient Integrator and
  the M\"obius Accelerated DWF Solver}

\ShortTitle{Improving DWF Simulations}

\author{\speaker{Hantao Yin} \\
  Department of Physics, Columbia University, New York, NY 10025,
  USA\\
  E-mail: \email{yinnht@phys.columbia.edu}}

\author{Robert D. Mawhinney\\
  Department of Physics, Columbia University, New York, NY 10025,
  USA\\
  E-mail: \email{rdm@physics.columbia.edu}}

\abstract{We have implemented a variant of the force gradient
  integrator proposed by Kennedy et.al. and are using it in our
  production 2+1 flavor DWF simulations with pion masses of 180 MeV in
  $(4.5 \textrm{fm})^3$ volumes. We find modest speed-ups ($\sim$
  20\%) from using the force gradient integrator, compared to our
  previously used Omelyan integrator. On other ensembles, primarily
  finite temperature 2+1 flavor DWF QCD, we have extensively tuned the
  Hasenbusch preconditioning masses and achieved speed-ups of
  2-3x. Here we have also switched to the force gradient integrator,
  but this change has not had any impact on the speed. We also report
  on an improved solver for DWF, which uses M\"obius fermions, with a
  smaller fifth dimension than the original DWF fermions, as an
  intermediate step in the generation of solutions of the Dirac
  equation. This approach cuts the number of effective Dirac
  applications by approximately a factor of 2 when the conjugate
  gradient iteration count is large.

}

\FullConference{XXIX International Symposium on Lattice Field Theory
  \\
                July 10 - 16 2011\\
                Squaw Valley, Lake Tahoe, California}

\begin{document}

\section{Introduction}

Ensemble generation constitutes a large portion of the computing time
in a lattice calculation, and in many cases is the most time consuming
part. Most lattice calculations use a variant of the Hybrid Monte
Carlo (HMC) algorithm to generate lattice configurations for later
measurements. As such, optimizing the HMC integration scheme is
important for many lattice calculations.

We make use of the Hasenbusch mass splitting method and the recently
devised force gradient integrator to optimize the molecular dynamics
evolution in our lattice computations. In particular we propose a very
efficient way to implement the force gradient integrator. The use of
this new integrator in combination with the Hasenbusch mass splitting
scheme has achieved a factor of 2 speed up in many of our practical
calculations.




Domain wall fermions (DWF) are widely used for many lattice QCD
calculations. To extend existing efforts on DWF physics, we present a
method that employs the similarity between the kernels of M\"obius
fermion and standard DWF to help accelerate the solution of the DWF
Dirac equations.

\section{Force Gradient Integrator Implementation}

As shown in \cite{FGI} the force gradient PQPQP integrator (FGI) can be
expressed as
\begin{equation}
  e^{\frac{1}{6}\tau \zeta_S}e^{\frac{1}{2}\tau \zeta_T}
  e^{\frac{2}{3}\tau \zeta_S-\frac{1}{72}
    \tau^3 \zeta_{\left\{S,\left\{S,T\right\}\right\}}}
  e^{\frac{1}{2}\tau \zeta_T}e^{\frac{1}{6}\tau \zeta_S}
\end{equation}

The key step for this integrator involves the following update
on the momentum field
\begin{equation}\label{fg_update}
  p_i\leftarrow 
  p_i -\frac{2}{3}\tau e_i(S)+\frac{1}{36}\tau^3 e^j(S)e_j e_i(S)
\end{equation}

$e_i$ can be seen as the generalization of the vector fields
$\partial/\partial x$ in the SU(3) Lie group space of link
variables. It follows the Leibniz rule and when acting on the gauge
field it is effectively equivalent to left multiplication by the
corresponding generator $e_i(U)=T_i U$.

To implement the second order derivative term $e^j(S) e_j e_i (S)$ in
FGI, an approach using the Taylor expansion can be used. Since $e_i$
and $e_j$ are interchangeable in this extra term, the following
approximation holds
\begin{equation}\label{fg_approx}
  \frac{2}{3}\tau e_i(S)-\frac{1}{36}\tau^3 e^j(S)e_j
  e_i(S)=\frac{2}{3}\tau \left(1-\frac{1}{24}\tau^2 F^j
  e_j\right)e_i(S)
  =\frac{2}{3}\tau e^{-\frac{1}{24}\tau^2 F^j
  e_j}e_i(S)+\mathcal{O}(\tau^5)
\end{equation}

Here $F^j=e^j(S)$ and it must be treated as independent of link
variables in the above equation($e_i$ acting on $F^j$ is defined to be
0 despite the fact that $F^j$ depends on the gauge field). Note that
since the FGI has a local error of $\mathcal{O}(\tau^5)$, the above
approximation does not change the size of the error.

Effectively we obtain from the last expression
\begin{equation}
  e^{-\frac{\tau^2}{24}F^j e_j}e_i(S) =e_i(S[U'])
\end{equation}
with
\begin{equation}
  U'=e^{-\frac{\tau^2}{24}F^j e_j}U=e^{-\frac{\tau^2}{24}F^j T_j}U
\end{equation}

As a consequence we implement the force gradient step using the
following two-step approach
\begin{enumerate}
\item (the force gradient step) Calculate $F^j=e^j(S[U])$. Then update
  the gauge field temporarily according to
  $U'=e^{-\frac{\tau^2}{24}F^j T_j}U$.
\item (the regular step) Calculate again $e_i(S[U'])$ using the
  updated gauge field. The corresponding force is added to the
  momentum field. Upon the completion of this step, restore the gauge
  field to $U$.
\end{enumerate}

The above implementation of the force gradient integrator is different
from \cite{FGI}, since the approximation in eq.(\ref{fg_approx}),
changes the $\mathcal{O}(\tau^5)$ error term. The resulting integrator
is still symmetric and symplectic.

There is an additional benefit from using this implementation. The two
steps described above are exactly the same except using different
gauge fields $U$ and $U'$, and $U'$ differs from $U$ by only a small
factor ($\sim \tau^2/24$). If a fermion action is involved, the
solution from the first step can be used as an initial guess for the
second step when solving the Dirac equations for the force terms. This
``force gradient forecasting'' reduces costs significantly. Unlike the
chronological solver, it is independent of the direction of the
molecular dynamics time and does not impact reversibility.

When testing the force gradient integrator on a $16^3\times 32\times
16$ lattice with 2+1 flavor dynamical DWF fermions (420MeV pion) using
quotient and rational quotient actions, we were able to raise the top
level step size from 1/4 as in the Omelyan to 1/3 for a unit
trajectory length. However, due to the extra cost in the force
gradient solve, the new integrator does not provide a significant
speed up over a finely tuned Omelyan integrator in the case we tested.
For larger $32^3$ volumes, we have seen a modest benefit from the FGI
and will test on $48^3$ volumes soon, where the benefit may be larger.

\section{Hasenbusch Mass Splitting}
%
%
%
%
%

The Hasenbusch factorization of the DWF fermion determinant entering
in our evolutions
\begin{equation}
  \det\left(\frac{D^\dag(m)D(m)}{D^\dag(1)D(1)}\right)=
  \det\left(\frac{D^\dag(m)D(m)}{D^\dag(m+\mu)D(m+\mu)}\right)
  \det\left(\frac{D^\dag(m+\mu)D(m+\mu)}{D^\dag(1)D(1)}\right)
\end{equation}
has been proposed in\cite{mass_split} to separate the contribution to
the fermion force from the light eigenmodes and heavier ones. The
inversion of $D^\dag(m)D(m)$ is more expensive than the introduced
operator $D^\dag(m+\mu)D(m+\mu)$. Urbach \textit{et al.}\cite{urbach}
proposed that by assigning a smaller force to the more expensive part
(via tuning $\mu$) and putting it on a coarser time scale by means of
a multiple level integrator, an acceleration of the molecular dynamics
evolution can be achieved. We have used this approach for a number of
years, but there are overheads associated with multiple time scale
integrators.

Here we propose an approach that uses the idea of mass splitting, but
without different time scales. Instead of putting the part with larger
force on a finer time scale, we split the force further via
introducing additional intermediate masses. Namely we introduce
\begin{equation}
  \det\left(\frac{D^\dag(m)D(m)}{D^\dag(1)D(1)}\right)=
  \prod^{k+1}_{i=1}\det\left(
  \frac{D^\dag(m+\mu_{i-1})D(m+\mu_{i-1})}{D^\dag(m+\mu_i)D(m+\mu_i)}
  \right)
\end{equation}
with $0=\mu_0<\mu_1<\cdots<\mu_{k+1}=1$. One advantage of this scheme
over \cite{urbach} is that all intermediate masses $\mu_i
(i=1,2,\cdots,k)$ can be tuned continuously, while any multiple
time-step integrator scheme only admits a $1:n$ ratio on the number of
force term evaluations, with integer $n$ being the number of
evaluations of the nested integrator per evaluation of the upper level
integrator.

The observed benefit of this scheme is that all Dirac equations
related to the force terms of the quotient actions need not be solved
very accurately, if enough intermediate masses are introduced and
tuned. We were generally able to solve the force term Dirac equations
only up to a residual of $10^{-6}$ instead of the previously used
$10^{-8}$, thus saving a considerable amount of time.

\FloatBarrier

\section{M\"obius Accelerated Domain Wall Fermion (MADWF)}

We start by considering the general Dirac equation, 
\begin{equation}\label{dirac_eq}
  D_{DW}(m)x=b, 
\end{equation}
with $D_{DW}$ being the standard domain wall fermion Dirac
operator. The goal is to use M\"obius fermions to construct an
approximate equation. Before solving equation (\ref{dirac_eq}), we
solve the approximated equation and use the solution as an initial
guess.

The primary tool to relate the standard domain wall fermion and the
M\"obius fermion is the domain wall - overlap
transformation\cite{MD_trans, borici_2004, borici_4d}. It relates the
M\"obius Dirac operator to an equivalent 4D overlap operator. By
applying $P^{-1} D_{DW}^{-1}(1)$ to both sides of equation
(\ref{dirac_eq})\footnote{$P$ is a simple matrix that shifts half of
  the spin component in the fermion vector by one in $s$ direction,
  its detailed form can be found in \cite{MD_trans}.}, we can
transform the 5D Dirac matrix into a diagonal form in the $s$
direction,

\begin{equation}\label{ov_eq}
  \left(
  \begin{array}{ccccc}
    D_{OV} & & & & \\
    S_2 S_3\cdots S_L(D_{OV}-1)d & 1 & & & \\
    S_3\cdots S_L(D_{OV}-1)d & & 1 & & \\
    \cdots & & & \cdots & \\
    S_L(D_{OV}-1)d & & & & 1 
  \end{array}
  \right)y = \left(
  \begin{array}{c}
    c_1 \\
    c_2 \\
    c_3 \\
    \cdots \\
    c_L
  \end{array}
  \right),
\end{equation}
with $y = (y_1, y_2, y_3, \cdots, y_L)^\mathrm{T} = P^{-1}x$, $L$ is
the size of the $s$ direction and
\begin{equation}\label{trans_5d_4d_source}
  c = (c_1, c_2, c_3, \cdots, c_L)^\mathrm{T}=P^{-1}D_{DW}(1)^{-1}b.
\end{equation}
The details, including the definition of symbols $S_2, \cdots, S_L$
and $d$, can be found in \cite{MD_trans}.

After the transformation the primary work is to solve the 4D equation
\begin{equation}\label{dwf_eq_4d}
  D_{OV}(m)y_1=c_1
\end{equation}
with an overlap-like operator $D_{OV}(m)$. The rest of the work
involves deriving the whole 5D solution from $y_1$, which can be
computed with little effort, as will be seen below.

We first address the question of solving the 4D equation
(\ref{dwf_eq_4d}). We notice that $D_{OV}(m)$ is an approximation to
the ideal overlap operator. In parallel we can also obtain a different
finite approximation $D'_{OV}(m)$ from any 5D M\"obius operator
$D'_{DW}(m)$. For our purpose we replace $D_{OV}(m)$ in equation
(\ref{dwf_eq_4d}), which comes from a standard domain wall fermion
operator, by $D'_{OV}(m)$ that comes from a M\"obius fermion
operator. The solution $y'_1$ to the following equation
\begin{equation}\label{mob_eq_4d}
  D'_{OV}(m)y'_1=c_1
\end{equation}
can be directly used to approximate $y_1$, and thus reconstruct a 5D
approximated solution to (\ref{dirac_eq}). 

To derive the entire 5D solution $y$ from $y_1$, we notice that for a
true solution $y=(y_1, y_2, y_3, \cdots, y_L)^\mathrm{T}$ we have
\begin{equation}
  y_k = c_k - S_k S_{k+1}\cdots S_L (D_{OV}-1)dy_1, \; k = 2,3,\cdots,L.
\end{equation}
This relation can also be used to construct the guessed solution once
$y_1'$, the approximation to $y_1$, is obtained. Namely,
\begin{equation}
  y'_k = c_k - S_k S_{k+1}\cdots S_L (D_{OV}-1)dy_1', \; k =
  2,3,\cdots,L, 
\end{equation}
where $y'_k$ represents the approximation to $y_k$. This can
be written in matrix form
\begin{equation}\label{recon}
\left(-D_{OV}y'_1,y'_2,y'_3,\cdots,y'_L \right)^\mathrm{T}
=P^{-1}D_{DW}(1)^{-1}D_{DW}(m)P\cdot\left(-y'_1,c_2,c_3,\cdots,c_L\right)^\mathrm{T}.
\end{equation}
Reconstructing the entire 5D guess in this way requires a
Pauli-Villars DWF solve.

There remains the question of solving the 4D M\"obius Dirac equation
(\ref{mob_eq_4d}). To do this we reverse the above process. The
detailed procedure can be found at \cite{MD_trans, borici_2004,
  borici_4d}. By using the domain wall - overlap transformation on the
4D equation (\ref{mob_eq_4d}) we can transform it back to the 5D form
\begin{equation}\label{mob_eq_5d}
  D'_{DW}(m)P y' = D'_{DW}(1)Pc'
\end{equation}
with $c'=(c_1, 0, 0, \cdots, 0)$ and $y'_1$ being the 4D component of
$y'$ on the $s=1$ hyperplane.

The algorithm then looks like
\begin{enumerate}
\item Construct $c_1$ from the original 5D source $b$, using equation
  (\ref{trans_5d_4d_source}).
\item Replace the 4D operator in equation (\ref{dwf_eq_4d}) by a
  4D operator from a M\"obius fermion operator with appropriately
  chosen parameters.
\item Transform the 4D M\"obius Dirac equation back to its 5D form
  (\ref{mob_eq_5d}) and solve the 5D equation, thus obtaining the 4D
  solution $y'_1$ to equation (\ref{mob_eq_4d}).
\item Use $y'_1$ as an approximation to the 4D domain wall Dirac
  equation (\ref{dwf_eq_4d}), and reconstruct the 5D (approximated)
  solution using (\ref{recon}).
\end{enumerate}

The above process requires two Pauli-Villars solves in standard DWF
((\ref{trans_5d_4d_source}) and (\ref{recon})) and one M\"obius solve
on a lattice with possibly smaller size in $s$ direction
(\ref{mob_eq_5d}). The gain over a direct solve comes from the fact
that the M\"obius equation has a potentially smaller $s$ dimension
size. It's clear that the approximated solution $y_1'$ must be
sufficiently close to the true solution $y_1$ to make this method
beneficial. This leads to the problem of choosing the best
parameters. The details of parameter tuning will be given later, in a
separate paper.

\section{Defect Correction and Production Use}
\label{sec:restart}

\begin{figure}[!htbp]
  \centering
  \input{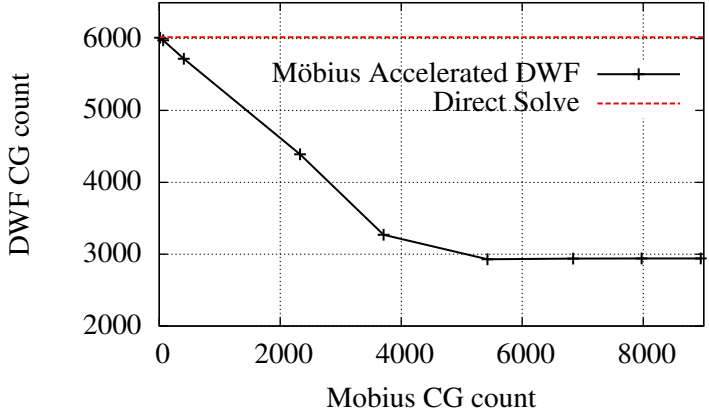}

  \caption{DWF CG count/M\"obius CG count correlation, 160MeV,
    $32^3\times 8$, with $L=32$, $L'=16$, $b=1.5$, $c=0.5$. A turning
    point is clearly presented in the curve, representing the best
    approximation can be obtained from solving the M\"obius Dirac
    equation.}
  \label{mob_rsd_acc}
\end{figure}


The M\"obius approximation to the standard DWF Dirac equation can only
achieve a certain accuracy (Figure (\ref{mob_rsd_acc})), as the two
operators are not exactly the same. A simple defect-correction method
can be used to achieve a higher accuracy in constructing the guessed
solution. In this approach, one takes the guessed solution and
calculates the residue, then constructs a second guess with the
residue as the new source vector using the M\"obius approach. The
second guess can be simply added to the original guess to improve
it. This method can be repeated if necessary.

\begin{table}[!htbp]
\centering
\begin{tabular}{|c||c||c|c|c|c|}
\hline
& Direct CG solve & \multicolumn{4}{c|}{M\"obius Accelerated DWF} \\
\hline
Operation & - & M\"obius equation(*3) & Pauli-Villars(*6) & others & total \\
\hline 
Op. Count & 11290*32=3.6e5 & 4.6e3*12 & 1.0e2*32 & - & 2.0e5 \\
\hline
time(s) & 2672 & 285 & 25 & 125 & 1138 \\
\hline
\end{tabular}
\caption{Cost comparison of MADWF CG solver with a regular (zero started)
  CG solver. $L=32$, $L'=12$. Data obtained from a 512 node partition
  on BG/P, both solve to $10^{-10}$. Note that there are 3
  Defect-correction steps, each includes 2 Pauli-Villars unit mass
  inversions. The factor 32 and 12 in the ``Op. Count'' row are due to
  different 5th dimension sizes.}
\label{full_example}
\end{table}

Table (\ref{full_example}) compares this method and the direct solve
approach on a $32^3\times 64$, $L=32$ lattice. The corresponding
M\"obius approximation has 5th dimension size $L'=12$. Since multiple
5th dimension sizes are used we count the number of 4D Wilson Dirac
operator applications (denoted by ``Op. count'' in the table). The
table shows that MADWF reduces the number of 4D Wilson Dirac operator
applications from 3.6e5 to 2.0e5, by a factor of 1.8. The MADWF solver
uses 42\% of wall clock time, when compared with a direct solve. Note
that while some gain is due to code optimization, most is due to the
fact that MADWF requires many fewer 4D Wilson Dirac operations.

We also note that the M\"obius parameters need only be tuned once for
a specific ensemble. Changing configuration or source vector has
little effect on the quality of the M\"obius approximation. Once
tuned, the same set of M\"obius parameters can be used to approximate
the DWF Dirac equation for the entire ensemble.


\FloatBarrier

\section{Conclusion}

We have presented a method to implement the force gradient
integrator. By testing the force gradient integrator on an existing
lattice we have shown that for small lattices the extra cost is high
enough so any benefit is small. However, due to its better scaling
behavior it's possible that the force gradient integrator performs
better than other second order integrators on larger lattices or with
longer trajectories.

We have explored the Hasenbusch mass splitting scheme and applied this
method to our practical calculation. We have found that extra quotient
actions introduced in this way allows the Dirac equations associated
with the force term evaluations to be solved less accurately, reducing
the cost of HMC considerably.

We have shown that the similarity between the 4D M\"obius fermion
Dirac operator and the Shamir domain wall fermion operator can be used
to accelerate solving the DWF Dirac equation. We have achieved a
factor of 2 speed up on a specific large lattices ($32^3\times 64$,
$L_s=32$) that we are using in practical calculations.

\section{Acknowledgment}

We thank A. Kennedy and M. Clark for notes regarding the force
gradient integrator and a prototype of a force gradient C++
implementation. We also thank our colleagues in the RBC and UKQCD
collaborations for discussions and tests on the program. This work is
done using the QCDOC supercomputers of the RBRC group, NYBlue at BNL
and the BG/P of the ALCF at Argonne. We use the MDWF package for
simulating the M\"obius fermions.

\end{document}